\documentclass[a4paper,twocolumn,10pt]{report}
\begin {document}
\twocolumn[%
\begin {center}
{\large \bf
Quantum gates using two-electron states of triple quantum dot
}\\
\ \\
H. Sasakura$^*$ and S. Muto\\
\newlength{\abswidth}
\setlength{\abswidth}{\linewidth}
\addtolength{\abswidth}{-2in}
%\begin {minipage}{\abswidth}
{\it 
Department of Applied Physics, Hokkaido University, N13 W8 Kitaku, Sapporo 060-8628, Japan}
\ \\
\ \\
\ \\
\begin {minipage}{\abswidth}
%Abstract
\ \ Quantum computation using electron spins in three coupled dot with different size is proposed. By using the energy selectivity of both photon assisted tunneling and spin rotation of electrons, logic gates are realized by static and rotational magnetic field and resonant optical pulses. Possibility of increasing the number of quantum bits using the energy selectivity is also discussed.\\ 
\end {minipage}
\end {center}
 \ \\ 
] 
%text
\ Since Shor's algorism was proposed, many studies have been carried out with a view to realize a quantum computer. The "computational" basis of a quantum computer is any superposition of two states, called "qubit" defined as $|\psi >=\cos \theta |0>+\sin \theta e^{i\alpha }|1>$, corresponding to classical(0,1) bit. Therefore, n-qubit can represent an arbitrary combination of $2^n$ informational states. It has been pointed out, with quantum mechanical effects of a parallel computation and interference, that an arbitrary physical system is efficiently simulated by a quantum computer.

Semiconductor quantum dots are attractive material for the quantum computation with its discrete electronic states and with its solid-state properties. Recently, technologies concerning the InAs/GaAs self-assembled quantum dots[1], [2] has been greatly developed especially with its closely stacking technique[3], [4], [5]. The InAs islands are found to align in the growth direction, as long as the GaAs barrier is thin enough below some 15nm, to form a column of InAs islands. We can control the tunneling barrier between quantum dots and also we can control effective height of a quantum dot by letting electronic wave function penetrating through nm-thick GaAs layers. Also, though electrons in quantum dots especially in the excited states are said to have short decoherence time, electron spins in quantum dots are believed to have much longer decoherence time. With these features in mind, we propose to use electron spins in three coupled quantum dots with different size, for the logic gates(one-bit rotation and controlled-NOT)[6] which can be operated by static and rotational (time-dependent) magnetic field and resonant optical pulses. With the different dot size, we can use the energy selectivity of both photon assisted tunneling and spin rotation of electrons.

{\bf Model} We consider only the electronic ground state of the quantum dot. We apply a static magnetic field {\bf B} $=(0,0,B)$ along the $z$ axis. {\it {Zeeman effect}} causes the spin splitting of the electronic ground state of a quantum dot with the energy splitting, $\Delta E$ (Fig. 1),\\
\begin {equation}
\displaystyle
\Delta E=E(\uparrow )-E(\downarrow )=\mu_{B} g B\ .
\end {equation}
Here, $\mu_{B}$ and $g$ denote the Bohr magneton and the effective g-factor of an electron.

According to the {\bf k}$\cdot ${\bf p} perturbation, the effective g-factor varies with the structure of the dot. For example, the g-factor of a sphere of "A" material embedded in "B" material is given by[7]\\
\ \\
$
\displaystyle
g(R)=g_{0}+[g_{A}(E)-g_{0}]w_{A}$
\begin {equation}
\displaystyle
+[g_{B}(E)-g_{0}]w_{B}
+[g_{B}(E)-g_{A}(E)]V(R)f^{2}(R)
\end {equation}
where $R$ is the dot radius, $g_{0}$ is Land$\grave {\mbox e}$ factor$(g_{0}\sim 2)$, $f(r)$ is the real part of the spinor envelope function, $g_{A}(E)(g_{B}(E))$ is the electron g-factor of "$A(B)$" material in the bulk, $V(R)$ is the volume of the dot and $w_{A}(w_{B})$ is $\int d{\bf r}f^{2}(r)$ taken over the "$A(B)$" volume. For instance, for a ${\mbox {GaAs/Al}}_{0.35}{\mbox {Ga}}_{0.65}\mbox {As}$ quantum dot, the effective g-factor of an electron was calculated to vary from -0.25 to 0.3 as the dot radius was varied from 5 nm to 15 nm.

{\bf Electronic States} Now, we consider three dots with different sizes under a static magnetic field ${\bf B}=(0,0,B)$. It causes the three electronic ground states splitted into six states, $|i>_{k}$ with energy,
\begin {equation}
\displaystyle
E_{i,k}=E_{k}+(-1)^{i}\mu _{B}g_{k}B/2 .
\end {equation}
Here, $k$ denotes the number of the three dots and $i(=0,1)$ indicates the up and down spin states of the dot. We can make differences of the energy of six states all different by using the size dependence of the effective g-factor of an electron (Fig. 2). In fact, the difference of the energy of six states are described as,
\begin {eqnarray}
\displaystyle
 & &\Delta E_{ijkl}=E_{i,k}-E_{j,l}\nonumber \\
& &
=E_{k}-E_{l}\nonumber \\
 & & \ +(-1)^{i}\mu _{B}g_{k}B/2-(-1)^{j}\mu _{B}g_{l}B/2\\
 & &\ \ \ (\{ i,j\} \in \{ 0,1\} \ ,\ \{ k,l\} \in \{ C,S,T\}) .\nonumber
\end {eqnarray}
We shine the (linearly polarized) resonant light which is resonant to the difference of the energy of six states. In the optical transition of the electron state by the resonant light, the spin state is conserved, $(i=j)$. Therefore, the operator, $C_{iikl}$, that indicates the optical transition between electronic states confined in the $k$-th quantum dot and the state in the $l$-th dot, is described as
\begin {equation}
\displaystyle
C_{iikl}\ : \ \ |i>_{k} \longleftrightarrow |i>_{l}\ .
\end {equation}
Here, we consider a three-quantum-dots system that has two-electrons. We only consider the coulomb energy, $U$ between two electrons in the same dot and neglect the electron-electron interaction otherwise. Also, we assume that $U$ is sufficiently larger than energy differences of single electron states. Therefore, we can neglect the possibilities of two electrons in a single dot.

{\bf One-bit Rotation} The qubit can be represented by the spin-1/2 state, $|\psi >=\cos \theta |\uparrow >+\sin \theta e^{i\alpha }|\downarrow >$, of an electron confined in a quantum dot[8]. We apply a static magnetic field {\bf B} $=(0,0,B)$ for the  zeeman splitting of the electronic ground state (Fig. 1). In addition, we apply a rotational magnetic field, resonant to the zeeman splitting
\begin {equation}
\displaystyle
B_{rot}=B_{1}\cos \omega _{B} t . \ \ \ \ (\hbar \omega _{B}=\mu _{B} g B)\ .
\end {equation}
An inversion of qubit states, $|0>$ and $|1>$, can be realized by the $\pi $ pulse of the rotational magnetic field,
\begin {eqnarray}
\displaystyle
R|\psi>&=&\nonumber R[\cos \theta |0>+\sin \theta e^{i\alpha }|1>]\\
\ &=&\cos \theta |1>+\sin \theta e^{i\alpha }|0>\ .
\end {eqnarray}
Here, $R$ denotes the unitary operator of the rotational magnetic field. If the n-qubit system is formed by n-dot with different sizes, we can rotate selectively the qubit with specific Zeeman Energy.

{\bf Controlled-NOT} In Fig. 2, we assign that the left "C" quantum dot as a {\it control dot}, the right "T" dot as a {\it target dot} and the center "S" dot as a {\it swap dot}. The controlled-NOT gate is realized by making the electron of "T" dot tunnel to the "S" dot depending on the spin state of the "C" dot, followed by application of magnetic $\pi $ pulse to the spin of the "T" dot. Here, we use the resonant transition by the optical pulse for the electron tunneling, i.e. photon assisted tunneling. The operator of the tunneling used are defined as,
\begin {equation}
\displaystyle
C_{1}=C_{11CS} \ \ , \ C_{2}=C_{11ST} \ \ , \ C_{3}=C_{00ST}
\end {equation}
\begin {eqnarray}
\displaystyle
C_{1}\ &:&\nonumber \ \ |1>_{C} \longleftrightarrow |1>_{S}\\
C_{2}\ &:& \ \ |1>_{S} \longleftrightarrow |1>_{T}\\
C_{3}\ &:&\nonumber \ \ |0>_{S} \longleftrightarrow |0>_{T}\ .
\end {eqnarray}
Also, we can invert selectly the target bit. This unitary operation is described by the rotational magnetic field operator, $R_{T}$,
\begin {equation}
\displaystyle
R_{T}\ : \ \ |0>_{T} \longleftrightarrow |1>_{T}\ .
\end {equation}

Details of the controlled-NOT operation are as follows. First, we assume the initial state where two-electrons are separately in the control dot and target dot, and swap dot has no electron;
\begin {eqnarray}
\displaystyle
(\alpha |0>_{C}+\beta |1>_{C})(\gamma |0>_{T}+\delta |1>_{T})\nonumber \\
\alpha ^{2}+\beta ^{2}=1\ ,\ \gamma ^{2}+\delta ^{2}=1\ .\nonumber
\end {eqnarray}
Second, we perform unitary operations, $\displaystyle CN\equiv C_{1}C_{2}C_{3}R_{T}C_{3}C_{2}C_{1}$, to the initial state. Fig. 3 shows the states realized by these operations. Consequently, states of qubits of control dot and target dot, transform as follows,\\
\ \\
$\displaystyle 
CN(\alpha |0>_{C}+\beta |1>_{C})(\gamma |0>_{T}+\delta |1>_{T})$\\
\ \\
\ \ \ \ \ \ \ $\displaystyle
=\alpha |0>_{C}(\gamma |0>_{T}+\delta |1>_{T})$
\begin {equation}\displaystyle
+\beta |1>_{C}(\gamma |1>_{T}+\delta |0>_{T})\ .
\end {equation}
This is the controlled-NOT operation, $\displaystyle |\epsilon _{C}>|\epsilon _{T}> \ \rightarrow |\epsilon _{C}>|\epsilon _{C}\otimes \epsilon _{T}>$(modulo2).

Here, we discuss the advantage of the energy selectivity of our model. With the energy selectivity, we can extend the model to many coupled dots with many qubits. Let us discuss the practical use of the selectivity by considering detuning effect on the error factor of the gate operation. First, we consider the simple two-level system. The flopping frequency is described as
\begin {equation}
\displaystyle
\tilde \Omega =\sqrt {\Omega ^{2}+(\omega _{0}-\omega)^{2}}
\end {equation}
where $(\omega _{0}-\omega )$ and $\Omega $ denote the detuning frequency and the Rabi frequency. We look for the condition that by an $\pi $ pulse of selected resonant transition, the 2n$\pi $ transition occurs to the detuned (non-selected) system, i.e. ,
\begin {equation}
\displaystyle
\pi /\Omega =2\pi n/\tilde {\Omega } \ \ \ \ (n=1,2,3,\cdots ) .
\end {equation}
Here, $\pi /\Omega $ is the "switching time"($T_{sw}$). For $n=1$, the detuning frequency, $\Delta $, is designed to be
\begin {equation}
\displaystyle
\Delta =\sqrt {3}\times \Omega =\sqrt {3}\times (\pi /T_{sw})\ \ \ \ \ \ (\Delta \equiv \omega _{0}-\omega) \ \ .
\end {equation}
The faster the switching time, the larger the detuning. If $T_{sw}$ is 10 ps, the detuning energy, $\hbar \Delta$, is 0.36 meV.\

Second, we consider the three-level system, or the "Vee" System[9]. We require larger detuning than the two-level system for the same maximum probability of "unintentional" transition. Fig. 4 shows the calculated change of 3 diagonal matrix elements of "Vee" System. The $\pi $ pulse ends at about 10 ps at which the "intentional" state "2" is almost fully populated. The "unintentional" state "3" is also populated but with a probability of less than 0.05. Here, the detuning frequency for 1 $\leftrightarrow$ 3 transition, $\Delta _{13}$, is set to be $2\sqrt {3}$ times the Rabi frequency of 1 $\leftrightarrow$ 2 transition, $\Omega _{12}$.

Let us go back to our model. First, the one-bit rotation gate operations at quantum dots can be seen as assembly of two-level transitions. The Rabi frequency is given by[10],
\begin {equation}
\displaystyle
\Omega _{Ri}=g_{i}\mu _{B} B_{1} / \hbar 
\end {equation}
where $g_{i}$ and $B_{1}$ denote the electron g-factor of $i$-th dot and the amplitude of the rotational magnetic field. The requirement for the smallest detuning is
\begin {equation}
\displaystyle
\Delta =|g_{i}-g_{j}|\mu _{B}B/\hbar \ge \sqrt {3}\Omega _{Ri} \ \ (i\neq j).
\end {equation}

There are two requirements for the optical $\pi $ pulse. First, the error caused by the difference of the spin (such as $C_{2}$ and $C_{3}$ in Fig. 2) can again be treated by the two-level system, and
\begin {equation}
\displaystyle
||\Delta E_{00kl}|-|\Delta E_{11kl}||\ge \sqrt {3}\hbar \Omega _{op}.
\end {equation}
Here $\Omega _{op}$ is the Rabi frequency of optical transitions which we design by the switching time as, $\pi /\Omega _{op}=T_{sw}$. The error caused by the choice of the quantum dot (such as $C_{1}$ and $C_{2}$ transitions) can be discussed by the "Vee" System. Therefore, the requirement is 
\begin {equation}
\displaystyle
||\Delta E_{iikl}|-|\Delta E_{iilm}||\ge 2\sqrt {3}\hbar \Omega _{op}
\end {equation}
where $i$ and ${k,l,m}$ denote the spin and quantum dots.

Now we consider an example of a simplified model of InAs/GaAs quantum dots[11]. We assume that the switching times of one operation, $R_{T}$, $C_{1}$, $C_{2}$ and $C_{3}$, are about 10 ps and static magnetic field, $B$ is 10 T. We assume the parabolic potential in the $xy$ plane and the square well in the $z$ axis. Then, the spinor wave function, $u_{s}=f_{z}\times f_{r}$, is\\
\[f_{z}= \left \{
\begin {array}{l@{\quad : \quad }l}
C_{z}\cos [kz] & |z| \leq d \\
C_{z}\cos [kd] \exp [-\kappa (|z|-d)] & |z|\ge d\\
\end {array} \right. \] 
\[f_{r}= 
\begin {array}{l@{\quad : \quad }l}
C_{lt}\exp [-\omega _{lt} r^{2}] & r^{2}=x^{2}+y^{2}\\
\end {array} \] 
$
\displaystyle
k=\sqrt { 2m_{A}(E)E/\hbar ^{2}}\ ,\ \omega _{lt}=\pi/8d_{lt}$\\
$\ \ \  ,\ \kappa =\sqrt {2m_{B}(E)(\Delta E_{c}-E)/ \hbar ^{2}}\ 
$\\
where $2d$ and $d_{lt}$ denote the well width and the lateral extension. The effective electron g-factor is described by
\begin {equation}
\displaystyle
g_{zz}=g_{0}+g_{zz}^{(2)}
\end {equation}
\begin {eqnarray}
\displaystyle
g^{(2)}_{zz} B_{z}&=&2\int d{\bf r} [u_{s}({\bf A}\times {\bf \nabla })_{z}
u_{s} ] (g(E)-g_{0}) \nonumber \\
&=&2\int d{\bf r}u_{s}^{2}(g(E)-g_0)B_{z} \nonumber
\end {eqnarray}
where {\bf A} denotes the vector potential. By using conditions (17)-(19) we could show that 9 quantum dots (5 qubits and 4 swap dots) can be controlled selectively with error rate of less than 0.1 (Fig. 5), (Table. I).
 This is not yet the optimized value. However, it is already large enough to demonstrate the potentiality of our model. So far we have assumed that the magnetic fields and optical pulses are specially uniform. If we can locally apply magnetic fields and optical pulses, the number of qubits could be infinite.

In summary, we have proposed the simple model of a quantum computer, using 6 levels of two-electrons confined in the three quantum dots. The basic ideas are 1) we neglect higher states of the electron confined by a quantum dot except the ground state, 2) The coulomb interaction is large enough to avoid the double occupancy of a quantum dot, 3) We use the electronic spin-1/2 state confined by a quantum dot as the "qubit", 4) We can operate the one-bit rotation to the qubit which we aim at using the dot size dependence of the effective g-factor of the quantum dot, 5) The controlled-NOT operation is realized by the simple combination of resonant transitions by linearly polarized light and the rotational magnetic field. Also, we have discussed the merit of energy selectivity of our model for its extension to many qubit system. 

The authors would like to thank professor R. Morita of Hokkaido University for helpful advises and discussions. The work is done by a Grant-in-Aid Scientific Research on the Priority Area "Spin Controlled Semiconductor Nanostructure" from the Ministry of Education, Science, Sports and Culture.
%REFERENCE
\begin {center}
{\large REFERENCE}
\end {center}
$[1]$ D. Leonard {\it et al}, Appl. Phys. Lett. {\bf 63}, 3203

   (1993).\\
$[2]$ J.-Y. Marzin {\it et al}, Phys. Rev. Lett. {\bf 73}, 716

   (1994).\\
$[3]$ Q. Xie, A. Madhukar, P. Chen and N. P. 

Kobayashi, Phys. Rev. Lett. {\bf 75} 2572 (1995).\\
$[4]$ G. S. Solomon, J. a. Trezza, A. F. Marshall and

 J. S. Harris Jr., Phys. Rev. Lett. {\bf 76}, 952 (1996).\\
$[5]$ Y. Sugiyama, Y. Nakata, K. Imamura, S. Muto 

and N. Yokoyama, Jpn. J. Appl. Phys. {\bf 35}, 1320

(1996).\\
$[6]$ A. Barenco {\it et al}, Phy. Rev. A. {\bf 52}, 3457 (1995).\\
$[7]$ A. A. Kiselev and E. L. Ivchenko, Phy. Rev. B. 

{\bf 58}, 16353 (1998).\\
$[8]$ D. Loss, and D. P. DiVicenzo, Phy. Rev. A. {\bf 57},

120 (1998).\\
$[9]$ B. W. Shore, {\it The theory of coherent atomic

 excitation}, Vol. 2.\\
$[10]$ C. P. Slichter, {\it Principles of Magnetic

 Resonance}.\\
$[11]$ C. Hermann {\it et al}, Phy. Rev. B. {\bf 15}. 823

 (1977).
\begin {center}
{\large CAPTIONS}
\end {center}
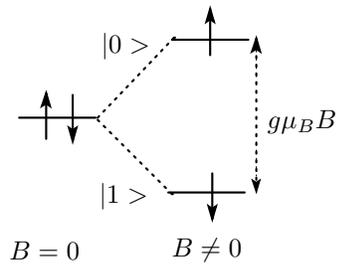
FIG. 1. Qubit using an electron spin.\\
\ \\
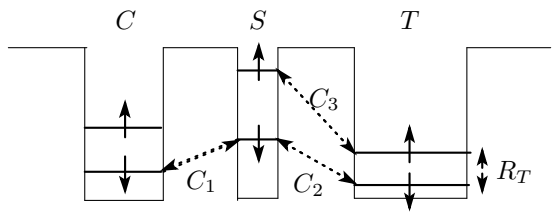
FIG. 2. Optical transitions (photon assisted

 tunneling), $C_{1}-C_{3}$, and spin rotation, $R_{T}$,

 used to realize the controlled-NOT. \\
\ \\
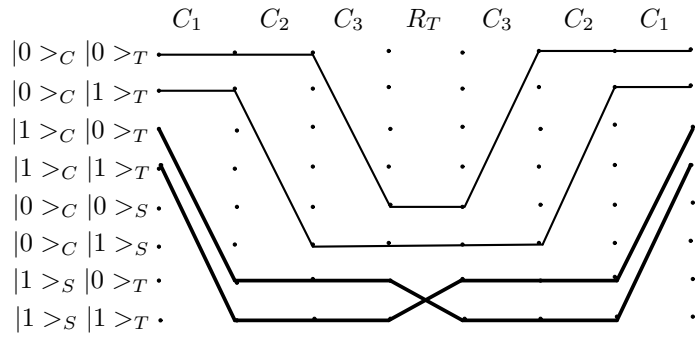
FIG. 3. 2 electron states realized during the

 controlled-NOT operation.\\
\ \\
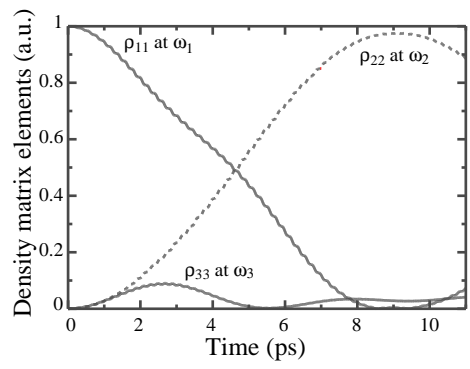
FIG. 4. Density matrix elements vs time for

 the "Vee" system. The $1\rightarrow 2$ transition is

 resonantly excited with $\hbar \omega _{2}-\hbar \omega _{1}=10$ meV. 

Transition has a detuning, $2\sqrt {3}\Omega _{12}=\hbar \omega _{3}$

$-\hbar \omega _{2}=0.72$ meV.\\
\ \\
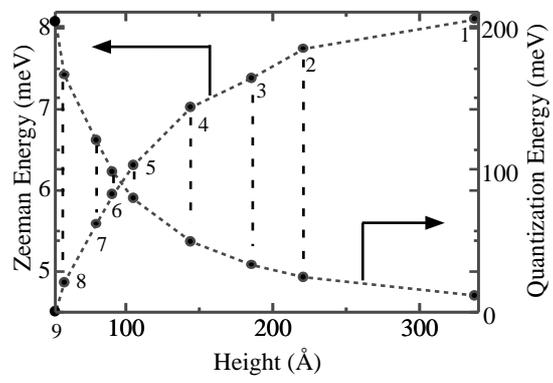
FIG. 5. The quantization energy, $E_{k}$, and

 the zeeman energy, $g_{k}\mu _{B}B$, for the dot heights

 chosen for the 9-dot InAs/GaAs system (B=10

 T).\\
\ \\
TABLE I. The resonant energies for optical

 transitions of 9 dots in Fig. 5, showing that the

 requirements (17) and (18) of the text are

 satisfied.
\ \\
\ \\
{\large E-mail address}\\
$^{*}$Electronic address: hirotaka@eng.hokudai.ac.jp\\
\ \\
\ \\
%Table
TABLE I
\ \\
\ \\
\begin {tabular}{ccc}
\hline \hline
\ \ \ k-l \ \ \ &\ \ \  $\Delta E_{00kl}$(meV)\ \ \  &\ \ \  $\Delta E_{11kl}$(meV)\ \ \ \\\hline
\ \ \ 1-2\ \ \  &\ \ \  13.2029 \ \ \ &\ \ \  12.4829\ \ \  \\
\ \ \ 2-3\ \ \  &\ \ \  \ 8.9224\ \ \  &\ \ \  \ 8.2024\ \ \  \\
\ \ \ 3-4\ \ \  &\ \ \  16.5682\ \ \  &\ \ \  15.8482\ \ \  \\
\ \ \ 4-5\ \ \  &\ \ \  14.9201\ \ \  &\ \ \  14.2001\ \ \  \\
\ \ \ 5-6\ \ \  &\ \ \  19.2130\ \ \  &\ \ \  18.4930\ \ \  \\
\ \ \ 6-7\ \ \  &\ \ \  22.1300\ \ \  &\ \ \  21.4100\ \ \  \\
\ \ \ 7-8\ \ \  &\ \ \  24.4699\ \ \  &\ \ \  23.7499\ \ \  \\
\ \ \ 8-9\ \ \  &\ \ \  37.3454\ \ \  &\ \ \  36.6254\ \ \  \\
\hline \hline
\end {tabular}
\end {document}